\documentclass[conference]{IEEEtran}
\IEEEoverridecommandlockouts
\usepackage{cite}
\usepackage{amsmath,amssymb,amsfonts}
\usepackage{graphicx}
\usepackage{textcomp}
\usepackage{subcaption}
\usepackage{threeparttable}
\usepackage{xcolor}
\usepackage{algorithm}
\usepackage{algpseudocode}
\def\BibTeX{{\rm B\kern-.05em{\sc i\kern-.025em b}\kern-.08em
    T\kern-.1667em\lower.7ex\hbox{E}\kern-.125emX}}
\begin{document}


\title{Fairness-Driven Optimization of RIS-Augmented 5G Networks for Seamless 3D UAV Connectivity Using DRL Algorithms\\

}

\author{\IEEEauthorblockN{Yu Tian$^1$, Ahmed Alhammadi$^1$,  Jiguang He$^1$, Aymen Fakhreddine$^{1}$, and Faouzi Bader$^{1}$}
\IEEEauthorblockA{$^1$Technology Innovation Institute, Abu Dhabi, United Arab Emirates}
\IEEEauthorblockA{E-mails: \{yu.tian, ahmed.alhammadi, jiguang.he, aymen.fakhreddine, carlos-faouzi.bader\}@tii.ae
}}

\maketitle

\begin{abstract}
In this paper, we study the problem of joint active and passive beamforming for reconfigurable intelligent surface (RIS)-assisted massive multiple-input multiple-output systems towards the extension of the wireless cellular coverage in 3D, where multiple RISs, each equipped with an array of passive elements, are deployed to assist a base station (BS) to simultaneously serve multiple unmanned aerial vehicles (UAVs) in the same time-frequency resource of 5G wireless communications. With a focus on ensuring fairness among UAVs, our objective is to maximize the minimum signal-to-interference-plus-noise ratio (SINR) at UAVs by jointly optimizing the transmit beamforming parameters at the BS and phase shift parameters at RISs. We propose two novel algorithms to address this problem. The first algorithm aims to mitigate interference by calculating the BS beamforming matrix through matrix inverse operations once the phase shift parameters are determined. The second one is based on the principle that one RIS element only serves one UAV and the phase shift parameter of this RIS element is optimally designed to compensate the phase offset caused by the propagation and fading. To obtain the optimal parameters, we utilize one state-of-the-art reinforcement learning algorithm, deep deterministic policy gradient, to solve these two optimization problems. Simulation results are provided to illustrate the effectiveness of our proposed solution and some insightful remarks are observed.
\end{abstract}


\section{Introduction}

Thanks to their convenient deployment, simple structure, and versatile landing capabilities, drones, also known as unmanned aerial vehicles (UAVs), have recently experienced significant growth. This expansion has given rise to a diverse array of applications, including but not limited to goods delivery, urban air taxis, remote surveillance, border control, agricultural and industrial monitoring, as well as disaster relief~\cite{Zhang2019,tian2022uav}. Despite the diverse nature of these applications, they share a common critical requirement: the necessity for robust three-dimensional (3D) wireless connectivity to facilitate the seamless exchange of real-time sensor data and control commands. The paramount concern across these applications is to ensure that the connectivity is both reliable and secure. This is imperative to prevent scenarios where prolonged control loss could potentially result in severe accidents.

Current widely deployed commercial fifth-generation (5G) base stations (BSs) are a suitable option for providing connectivity to UAVs. However, these BSs are specifically designed to offer two-dimensional (2D) coverage for ground users, resulting in weakened signal strength in the sky. Addressing the need for ubiquitous 3D coverage and global connectivity is a key focus for future wireless networks, i.e., beyond 5G and sixth generation (6G)~\cite{Zhang2019_VTM}. To overcome the absence of dedicated infrastructure for airborne UAVs, solutions that minimize the need for significant hardware upgrades or additional BS deployments by mobile network operators are preferable. Emphasizing reliance on existing 5G cellular networks and their enhancement to meet quality-of-service (QoS) requirements for UAV communications is essential. The authors of \cite{he2023unleashing} explore the feasibility of utilizing reconfigurable intelligent surfaces (RISs) to extend 3D coverage in cellular networks for aerial users. Their study indicates that deploying RISs can yield impressive gains in terms of achievable rate, outperforming traditional RIS-free 5G networks. However, the optimal power allocation and phase shift configuration for multi-RIS-augmented 3D coverage in multi-user cellular networks remain unexplored.


In the realm of recent research on RIS-augmented wireless communication systems, the predominant focus has been on two primary objectives for power allocation and RIS configuration optimization: maximizing the sum data rate \cite{huang2020reconfigurable, saglam2022deep,chen2021joint,xu2021ris,george2022deep} and maximizing the minimum signal-to-interference-plus-noise ratio (SINR) \cite{li2019joint,lu2021aerial}. The former objective centers around maximizing the overall throughput of the system, prioritizing system-wide efficiency. In contrast, the latter emphasizes the maximization of user fairness, with a focus on equitable resource distribution among users. As highlighted in \cite{zhou2023survey}, the existing literature predominantly employs deep reinforcement learning (DRL) to address throughput optimization challenges, while there is a noticeable scarcity of studies utilizing this approach to specifically tackle user fairness concerns. 
Capitalizing on the demonstrated success of DRL in addressing NP-hard non-convex optimization problems, our paper delves into the user fairness within a 3D multi-RIS-augmented multiple-input multiple-output (MIMO) system. Our objective is to maximize the minimum SINR through the joint optimization of transmit beamforming vectors and phase shift parameters. To achieve this and alleviate the complexities associated with the massive search space, we introduce the innovative ``One RIS Element Serving One UAV" (ORESOU) approach and the interference canceling (IC) method. Leveraging cutting-edge DRL algorithm, deep deterministic policy gradient (DDPG), we implement and evaluate the performance of our proposed algorithms.


\section{System Model}
\begin{figure}[htb!]
	\centering
\includegraphics[width=0.9\linewidth]{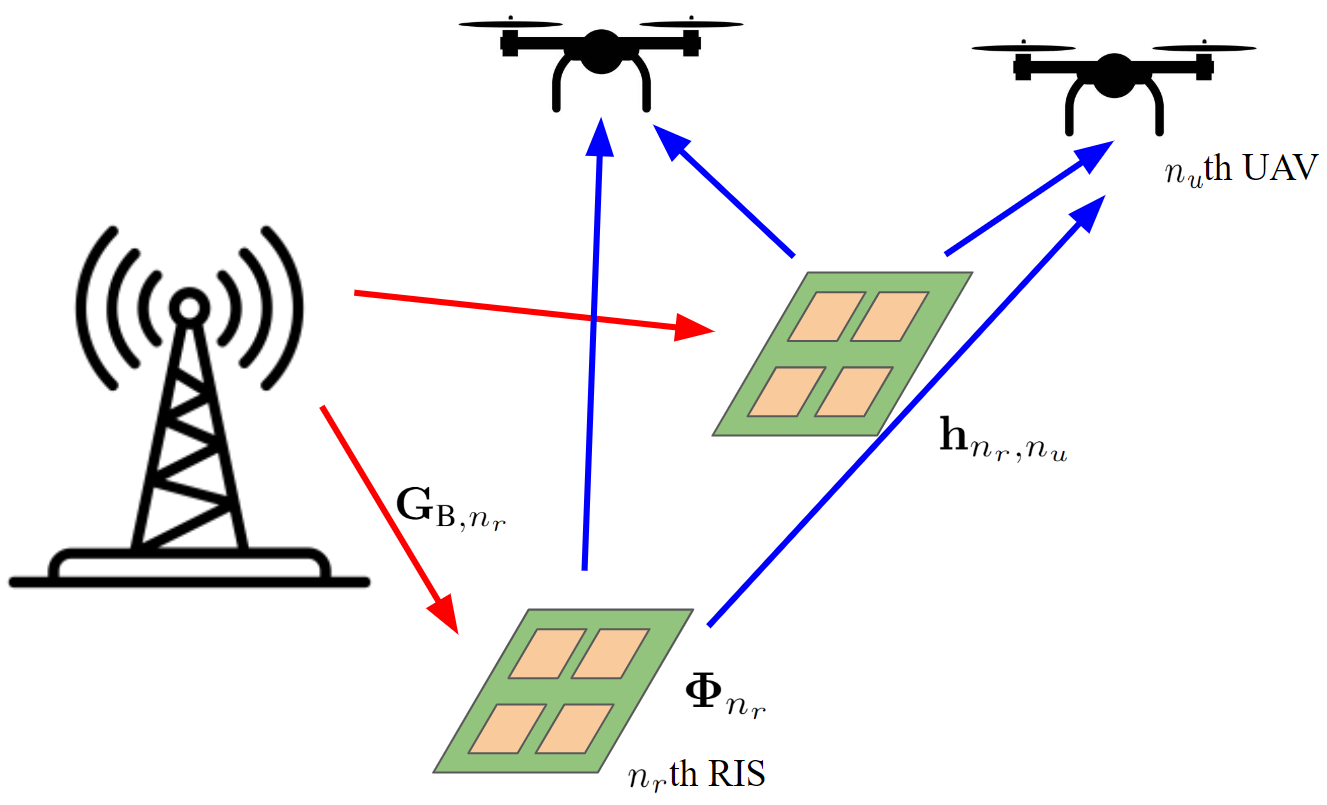}
	\caption{3D connectivity enabled by multiple RISs along with a legacy 5G BS.}
		\label{System_Model}
\end{figure}

The configuration of the system model is depicted in Fig.~\ref{System_Model}. Our focus is on a massive MIMO system aided by multiple RISs. The system comprises a single base station (BS) equipped with $N_\text{B}$ vertical antenna elements, $N_\text{R}$ RISs, and $N_\text{U}$ UAVs, each equipped with a single omnidirectional antenna.
Each RIS consists of $N=N_{x}\times N_{y}$ reflecting meta-atoms, where $N_{x}$ and $N_{y}$ denote the number of RIS meta-atoms along the $x$ and $y$ axes, respectively. For the sake of simplifying our analytical approach, we make the assumption that the RIS is square, possessing an equal number of meta-atoms along both the $x$ and $y$ axes, i.e., $N_x=N_y$.

We assume that the BS exclusively serves terrestrial users, and its beams are down-tilted. Simultaneously, the UAVs operate at a substantial altitude, beyond the terrestrial coverage area. Consequently, direct links between the BS and UAVs are nonexistent. In this context, RISs can be strategically positioned on the ground, walls, or atop buildings to enhance communication between the BS and UAVs by intelligently reflecting signals from the BS. All the RISs are connected and receive a feedback channel from the BS.

Let $\mathbf{G}_{\text{B},n_r}\in \mathbb{C}^{{N_\text{B}}\times N}$ denote the channel from the BS to the $n_r$-th RIS, and $\mathbf{h}_{n_r,n_u}\in \mathbb{C}^{1\times N}$ represent the channel from the $n_r$-th RIS to the $n_u$-th UAV. Typically, RISs are positioned in locations with a direct link to the BS to ensure high signal quality. Nevertheless, the presence of nearby objects inevitably leads to pervasive reflection, inducing small-scale fading. Therefore, we model the BS-RIS channel by employing the Rician fading which is a summation of weighted line-of-sight (LoS) and non-line-of-sight (NLoS) paths, expressed mathematically as:
\begin{align}\mathbf{G}_{\text{B},n_r}=\underbrace{\sqrt{\frac{1}{\kappa_1+1}}\widetilde{\mathbf{G}}_{n_r}}_\text{NLoS}+\underbrace{\sqrt{\frac{\kappa_1}{\kappa_1+1}}\overline{\mathbf{G}}_{n_r}}_\text{LoS},
\end{align}
where $\widetilde{\mathbf{G}}_{n_r}$ represents the NLoS component, $\overline{\mathbf{G}}_{n_r}$ is the LoS component, and $\kappa_1$ denotes the power ratio between $\overline{\mathbf{G}}_{n_r}$ and $\widetilde{\mathbf{G}}_{n_r}$. Each entry of $\widetilde{\mathbf{G}}_{n_r}$ follows a complex Gaussian distribution $\mathcal{CN}(0,1)$. $\overline{\mathbf{G}}_{n_r}$ is expressed with respect to the steering vector for the square RIS with ideal isotropic elements and the steering vector of the BS. These vectors depend on the azimuth and elevation angles of arrival and departure for the signals. Specifically, 
\begin{align}\overline{\mathbf{G}}_{n_r}=\boldsymbol{\alpha}_\text{R} (\theta^r_{\text{R},n_r}, \phi^r_{\text{R},n_r}) 
\boldsymbol{\alpha}_\text{B}^\mathsf{H} (\phi^t_{\text{B},n_r}- \beta)\exp{(-j2\pi f_c\tau_{\text{B},n_r})}, 
\end{align}
where $\theta^r_{\text{R},n_r} \in [0,\pi]$, $\phi^r_{\text{R},n_r} \in [-\pi/2,\pi/2]$, and $\phi^t_{\text{B},n_r} \in [-\pi/2,\pi/2]$ represent the azimuth, elevation angle of arrival (AoA), and elevation angle of departure (AoD), respectively, $\beta$ denotes the down-tilted orientation at the BS. Here, we assume that all the communications share the same time-frequency resources, $f_c$ denotes the carrier frequency, and $\tau_{\text{B},n_r}$ is the propagation delay from the BS to the $n_r$-th RIS. $\boldsymbol{\alpha}_\text{R} (\theta^r_{\text{R},n_r}, \phi^r_{\text{R},n_r})$ and $\boldsymbol{\alpha}_\text{B}(\phi^t_{\text{B},n_r})$ are the normalized array response steering vectors, defined as  
\begin{align}\label{alpha}
   & \boldsymbol{\alpha}_\text{R} (\theta^r_{\text{R},n_r}, \phi^r_{\text{R},n_r}) \notag\\
   &\triangleq \frac{1}{\sqrt{N}}\Bigg[1, \exp\left(j \frac{2\pi  d_{x}}{\lambda}  \cos(\theta^r_{\text{R},n_r}) \sin(\phi^r_{\text{R},n_r})\right), \nonumber\\
    &\;\;\; \cdots, \exp\left(j \frac{2\pi d_{x}}{\lambda} (N_{x} -1) \cos(\theta^r_{\text{R},n_r})\sin(\phi^r_{\text{R},n_r})\right) \Bigg]^{\mathsf{T}}  \nonumber \\& \otimes\Bigg[1, \exp\left(j \frac{2\pi  d_{y}}{\lambda}  \sin(\theta^r_{\text{R},n_r}) \sin(\phi^r_{\text{R},n_r})\right), \nonumber\\
    &\;\;\; \cdots, \exp\left(j \frac{2\pi d_{y}}{\lambda} (N_{y} -1) \sin(\theta^r_{\text{R},n_r})\sin(\phi^r_{\text{R},n_r})\right) \Bigg]^{\mathsf{T}},\\
       &\boldsymbol{\alpha}_\text{B}(\phi^t_{\text{B},n_r}- \beta) \triangleq \frac{1}{\sqrt{N_\text{B}}}\Bigg[1, \exp\left(j \frac{2\pi d_{z}}{\lambda}  \cos(\phi^t_{\text{B},n_r}- \beta) \right), \nonumber\\
    &\;\;\; \cdots, \exp\left(j \frac{2\pi d_{z}}{\lambda} (N_\text{B} -1) \cos(\phi^t_{\text{B},n_r}- \beta)\right) \Bigg]^{\mathsf{T}},     
 \end{align}
 where $d_x$, $d_y$, and $d_z$ represent the inter-element  spacing across $x$, $y$, and $z$ axes, respectively. $\otimes$ denotes the Kronecker product.
 We can easily see that $\|\boldsymbol{\alpha}_\text{R} (\theta^r_{\text{R},n_r}, \phi^r_{\text{R},n_r})\|_2 = \|\boldsymbol{\alpha}_\text{B}(\phi^t_{\text{B},n_r}- \beta)\|_2 = 1$. 

Meanwhile, the channels between the $n_r$-th RISs and the $n_u$-th UAVs are also modeled by Rician fading, expressed as 
\begin{align}
    \mathbf{h}_{n_r,n_u}=\sqrt{\frac{\kappa_2}{\kappa_2+1}}\overline{\mathbf{h}}_{n_r,n_u}+\sqrt{\frac{1}{\kappa_2+1}}\widetilde{\mathbf{h}}_{n_r,n_u},
\end{align}
where each element of $\widetilde{\mathbf{h}}_{n_r,n_u}$ follows $\mathcal{CN}(0,1)$ and $\overline{\mathbf{h}}_{n_r,n_u}=\boldsymbol{\alpha}_\text{R}^\mathsf{H} (\theta^t_{n_r,n_u},\phi^t_{n_r,n_u})\times\exp{(-j2\pi f_c\tau_{n_r,n_u})}$ with $\theta^t_{n_r,n_u}$ and $\phi^t_{n_r,n_u}$ denoting the azimuth and elevation AoDs from the RIS, and $\tau_{n_r,n_u}$ representing the propagation delay between the $n_r$-th RIS and the $n_u$-th UAV.

Hence, the channel from the BS to the $n_u$-th UAV through the $N_{\text{R}}$ RISs can be expressed as  
\begin{align}
    \boldsymbol{\omega}_{\text{B},n_u} = \sum_{n_r=1}^{N_{\text{R}}} \mathbf{h}_{n_r,n_u}  \mathbf{\Phi}_{n_r} \mathbf{G}_{\text{B},n_r}
\end{align}
where 
\begin{align}
    \mathbf{\Phi}_{n_r} = \mathrm{diag}(\exp{(\theta_{n_r,1})},...,\exp{(\theta_{n_r,N})})
\end{align} 
is the diagonal phase control matrix of the $n_r$-th RIS and $\mathrm{diag}(\mathbf{a})$ denotes a diagonal matrix with the entries of $\mathbf{a}$ on its diagonal. 

Furthermore, the received signal at the $n_u$-th UAV comprises the desired signal, interference signal from the other UAVs, and the noise, expressed as 
\begin{align}
    y_{n_u} =  \boldsymbol{\omega}_{\text{B},n_u} \left(\sqrt{P_{n_u}} \mathbf{f}_{n_u} s_{n_u}+\sum_{i=1,i\neq n_u}^{N_{\text{U}}} \sqrt{P_i} \mathbf{f}_i s_i\right) + z_{n_u},
\end{align}
where $P_{n_u}$,  $\mathbf{f}_{n_u}\in \mathbb{C} ^{N_{\text{B}}}$, and $s_{n_u}\sim \mathcal{CN}(0,1)$ are respectively the transmit power, the beamforming vector, and the transmitted signal symbol for the ${n_u}$-th UAV. $z_{n_u}$ is the additive complex Gaussian noise with zero mean and variance of $\sigma_z^2$, i.e., $\mathcal{CN}(0,\sigma_z^2)$.

Finally, the SINR of the $n_u$-th UAV is expressed as 
\begin{align}\label{sinr}
    \gamma_{n_u}=\frac{ P_{n_u} |\boldsymbol{\omega}_{\text{B},n_u}  \mathbf{f}_{n_u}|^2}{\sum_{i \neq n_u } P_{i} |\boldsymbol{\omega}_{\text{B},n_u}  \mathbf{f}_{i}|^2+ \sigma_z^2},
\end{align}
and the sum rate of all the $N_{\text{U}}$ UAVs is presented as
\begin{align}
    R = \sum_{n_u=1}^{N_{\text{U}}} \log_2 (1+\gamma_{n_u}).
\end{align}

\section{Problem Formulation}

Most of the existing literature \cite{huang2020reconfigurable, saglam2022deep,chen2021joint,xu2021ris,george2022deep} focus on maximizing the sum rate of all users, potentially leading to fairness concerns as illustrated in Fig.~\ref{obj_sr}. This objective function tends to allocate more power to users with superior channel gains while less power to those with inferior channel gains. Motivated by~\cite{li2019joint}, we opt for the objective function of maximizing the minimal SINR.
The optimization problem can be expressed as
\begin{align}\label{maxminsinr}
&\max\limits_{\mathbf{\Phi},\mathbf{F},\mathbf{P}}\,\, \min\limits_{n_u}\gamma_{n_u}\\
&\text{s.t.}\begin{cases}
||\mathbf{P}||_1\leq P_{\max}\\
\mathrm{diag}(\mathbf{F}^\mathsf{H}\mathbf{F})=\mathbf{I}_{N_{\text{U}}}\\
|\mathbf{\Phi}|=\mathbf{I}_{NN_{\text{R}}}\\\end{cases}\label{1conditions},
\end{align}
where $\mathbf{P}=\mathrm{diag}(P_1,...,P_{N_{\text{U}}})\in \mathbb{R}^{N_{\text{U}}\times N_{\text{U}}}$, $\mathbf{F}=[\mathbf{f}_1,...,\mathbf{f}_{N_{\text{U}}}]\in \mathbb{C}^{N_{\text{B}}\times N_{\text{U}}}$, $\mathbf{\Phi}=\mathrm{diag}(\mathbf{\Phi}_1,...,\mathbf{\Phi}_{N_{\text{R}}})\in \mathbb{C}^{NN_{\text{R}}\times NN_{\text{R}}}$, and $P_{\max}$ represents the maximum transmit power at the BS. 

The optimization problem \eqref{maxminsinr} is non-convex, involving three variable matrices. Presently, many researchers resort to reinforcement learning (RL) to jointly optimize all the three variables simultaneously, which suffers from a vast search space. In this paper, we propose two algorithms to tackle this issue. 

\subsection{Interference Cancellation (IC) Algorithm}\label{ICAlo}

The overarching objective of maximizing the minimum SINR in \eqref{maxminsinr} is to simultaneously maximize the power of the received desired signal and minimize the interference power. The ideal outcome is characterized by zero interference power, with all users sharing the same maximized received power. Consequently, the objective function can be formulated as:

\begin{align}\label{interfcancel}
&\max\limits_{\mathbf{\Phi},\mathbf{F},\mathbf{P}}\,\, \alpha\\
&\text{s.t.}\begin{cases}\mathbf{H} \mathbf{\Phi}  \mathbf{G} \mathbf{F}\mathbf{P} = \alpha \mathbf{I}_{N_{\text{U}}}\\
||\mathbf{P}||_1\leq P_{\max}\\
\mathrm{diag}(\mathbf{F}^\mathsf{H}\mathbf{F})=\mathbf{I}_{N_{\text{U}}}\\
|\mathbf{\Phi}|=\mathbf{I}_{NN_{\text{R}}}\\\end{cases}\label{conditions}
\end{align}
where $\mathbf{G}=[\mathbf{G}_{\text{B},1},...,\mathbf{G}_{\text{B},N_{\text{R}}}]\in \mathbb{C}^{NN_{\text{R}}\times N_{\text{B}} }$ and $\mathbf{H} = [[\mathbf{h}_{1,1},...,\mathbf{h}_{NN_{\text{R}},1}];...;[\mathbf{h}_{1,N_{\text{U}}},...,\mathbf{h}_{NN_{\text{R}},N_{\text{U}}}]]\in \mathbb{C}^{N_{\text{U}} \times NN_{\text{R}} }$. 

Because $\mathbf{\Phi}$ and $\mathbf{P}$ are diagonal, $\mathbf{H} \mathbf{\Phi}  \mathbf{G} \in \mathbb{C}^{N_{\text{U}}\times N_{\text{B}}}$ and $\mathbf{F}\mathbf{P}\in \mathbb{C}^{N_{\text{B}}\times N_{\text{U}}}$ should be reciprocal to meet the first condition in \eqref{1conditions}. $\mathbf{FP}$ can be combined together to one variable $\mathbf{\hat{F}}=\mathbf{FP}\in \mathbb{C}^{N_{\text{B}}\times N_{\text{U}}}$ with condition $\mathrm{trace}(\mathbf{\hat{F}}^\mathsf{H}\mathbf{\hat{F}})\leq P_{\max}$. Once $\mathbf{\Phi}$ is obtained, $\mathbf{\hat{F}}$ can be obtained by the pseudo inverse of $\mathbf{H} \mathbf{\Phi}  \mathbf{G}$, $\mathbf{\hat{F}}=\hat{P}\frac{\mathbf{\widetilde{F}}}{\mathrm{trace}(\mathbf{\widetilde{F}}^\mathsf{H}\mathbf{\widetilde{F}})}$ with $\mathbf{\widetilde{F}}= \mathrm{pinv}(\mathbf{H} \mathbf{\Phi}  \mathbf{G})$ and $\hat{P}\leq P_{\max}$. In this way, we can get 
\begin{align}
    \mathbf{H} \mathbf{\Phi}  \mathbf{G} \mathbf{\hat{F}}=\hat{P}\frac{\mathbf{H} \mathbf{\Phi}  \mathbf{G}\mathbf{\widetilde{F}}}{\mathrm{trace}(\mathbf{\widetilde{F}}^\mathsf{H}\mathbf{\widetilde{F}})}=\hat{P}\frac{\mathbf{I}_{N_{\text{U}}}}{\mathrm{trace}(\mathbf{\widetilde{F}}^\mathsf{H}\mathbf{\widetilde{F}})}.
\end{align}
Note that $\mathbf{H} \mathbf{\Phi}  \mathbf{G}$ and $\mathbf{\hat{F}}$ are non-square matrices. They should be full-rank matrices to satisfy the pseudo inverse requirements. Namely, $\mathrm{rank}(\mathbf{H} \mathbf{\Phi}  \mathbf{G})=\mathrm{rank}(\mathbf{\hat{F}})=\min{(N_{\text{U}},N_{\text{B}})}$.

To maximize above equation, $\hat{P}$ should be equal to $P_{\max}$, that is, $\hat{P}=P_{\max}$. Finally, we can get 
\begin{align}\label{finalfhat}
    \mathbf{\hat{F}} = P_{\max}\frac{\mathbf{\widetilde{F}}}{\mathrm{trace}(\mathbf{\widetilde{F}}^\mathsf{H}\mathbf{\widetilde{F}})}.
\end{align}

 Because the interference is canceled, the SINR of the $n_u$-th UAV in \eqref{sinr} becomes SNR, that is, $\gamma_{n_u}={ P_{n_u} |\boldsymbol{\omega}_{\text{B},n_u}  \mathbf{f}_{n_u}|^2}/{ \sigma_z^2}$. 
In this context, noise power plays a vital role in minimum SINR optimization. 
\subsection{One-RIS-Element-Serving-One-UAV (ORESOU) Algorithm}\label{ORESOUAlo}

$\mathbf{H} \mathbf{\Phi}  \mathbf{G}$ and $\mathbf{\hat{F}}$  may not always satisfy the full-rank requirement. To address this issue, we propose the ORESOU algorithm with an assumption that each RIS element serves only one UAV. The phase shift of this element is adjusted to compensate for the phase offset induced by the propagation and fading from the BS to the RIS element and from the RIS element to the target UAV. In this context, we can substitute the continuous $\mathbf{\Phi}$ with an association matrix $\mathbf{\overline{\Phi}}\in\mathbb{B}^{NN_{\text{R}}\times N_{\text{U}}}$ that links all the RIS elements and all the UAVs. $\mathbf{\overline{\Phi}}$ possesses the property that its elements take only two values from $\mathbb{B} = \{0,1\}$ and the sum of each row equals~1.

\section{Deep Reinforcement Learning Framework}
In this section, we provide a brief overview of DRL, laying the foundation for the proposed joint design of BS transmit beamforming and RIS phase shifts.

\subsection{Overview of RL}
RL is a branch of machine learning that deals with how agents should take actions in an environment to maximize some notion of cumulative reward \cite{huang2020reconfigurable}. In a typical RL, the agent incrementally refines its decision-making capabilities through ongoing interactions with the environment. During each interaction, the agent applies actions to the environment, observing the instant rewards and tracking state transitions  of the environment.

The primary objective of RL is to maximize the cumulative sum of rewards, as quantified by the state-action value functions. Given a state, RL can determine the optimal action based on the trained policy. In traditional RL algorithms, the state-action value function is computed and updated using the Bellman equation. To manage discrete states and actions, all corresponding values are stored in a Q-table. However, when dealing with continuous actions, such as beamforming and phase shift matrices in our problem, the Q-table becomes impractical. In such cases, a deep neural network is employed as a replacement for the Q-table, allowing for a more flexible and scalable representation of the state-action space which facilitates the creation of the DRL framework.

\subsection{Actor-Critic Framework}

The RL actor-critic framework is a hybrid architecture that combines a policy-based (actor) and a value-based (critic) approach. Actor defines the agent's policy that determines the probability distribution of actions for a given state, while critic evaluates the value of these actions by estimating a state-action value function. Actor evaluates the value of these actions based on the information from the critic's dominance information to update it, guiding the selection of actions that lead to higher than expected rewards. At the same time, critic is trained to minimise the time-difference error between the estimated and actual rewards, providing a benchmark for the actor's updates. This cooperative learning strategy improves stability and efficiency compared to purely policy-based or value-based approaches. The actor-critic framework, typically implemented using deep neural networks, is suitable for tasks with a continuous action space and has been successful in a variety of applications, including robot control and game play. Actor-critic frameworks implemented via deep neural networks (DNNs) perform well in continuous action space scenarios, overcoming the limitations of traditional Q-table approaches.

\subsection{Deep Deterministic Policy Gradient (DDPG)}

DDPG is a prominent off-policy actor-critic algorithm designed to solve RL problems in continuous action spaces. DDPG is characterised by the learning of deterministic policies, which allows for direct mapping from states to actions, which is particularly beneficial in scenarios with continuous and high-dimensional action spaces \cite{haarnoja2018soft}. The algorithm maintains a network of actors, which are responsible for policy approximation, and a network of critics, which evaluate the expected payoffs based on the actions chosen by the actors. DDPG employs experience replay, a mechanism for storing and randomly sampling past experience, which enhances the stability and efficiency of the learning by breaking down correlations between consecutive samples. In addition, the integration of a target network that periodically updates slower copies of the main network helps to achieve a more stable training process by providing consistent target values during learning updates.

The DDPG process involves iteratively updating actor and critic networks through a combination of policy gradient ascent and temporal difference learning. The actor is trained to maximise the expected return from the critic's perspective, and the critic is updated to minimise the mean-squared time-difference error between its predicted Q-value and the target Q-value obtained from the target network. This process is guided by the deterministic policy gradient theorem, which drives the actor's updates in the direction of maximising the expected return. The algorithm also explores by adding noise to the selection of actions, prompting a balance between exploration and exploitation. Overall, DDPG provides an efficient and stable solution to the continuous control problem and works in line with the principles of strategy optimisation and value estimation in the actor-critic framework.

\begin{algorithm}[htb!]
\caption{DDPG for IC and ORESOU algorithms}\label{ddpgic}
 \textbf{Input: $\mathbf{G}$, $\mathbf{H}$} \\
 \textbf{Output:} {Q-value function,  $\mathbf{\Phi}$ for IC,  $\{\mathbf{\overline{\Phi}},\hat{\mathbf{F}}\}$ for ORESOU} 
 \textbf{Initialization:} {Critic, Actor, and target networks, replay buffer}
\begin{algorithmic}[1]
\For{each episode}
\State Initialize $\mathbf{G}$ and $\mathbf{H}$, and receive first state
\For{each time step}
\State Obtain action ($\mathbf{\Phi}$  or  $\{\mathbf{\overline{\Phi}},\hat{\mathbf{F}}\}$) from Actor 
\State {Observe new state given action and noise, calculate best $\mathbf{\hat{F}}$ for IC and best $\mathbf{\Phi}$ for ORESOU }
\State Obtain minimum SINR
\State {Store state, action, new state, and minimum SINR to replay buffer}
\State Obtain Q-value from Critic
\State Sample random mini-batches from replay buffer
\State Update Critic by temporal difference learning
\State Update Actor by policy gradient ascent
\State Periodically slowly update target networks
\EndFor
\EndFor
\end{algorithmic}
\end{algorithm}
\subsection{DDPG for IC and ORESOU}
The implementation details of DDPG for IC and ORESOU are described in Algorithm \ref{ddpgic}. Both the IC and ORESOU adhere to the same reward and environmental setup.  The reward is defined as the minimum SINR, emphasizing the optimization objective. The environmental setup encompasses the spatial coordinates and channel information associated with the BS, RISs, and UAVs. 

The primary distinction lies in the action space. As outlined in Section \ref{ICAlo}, IC leverages DDPG to explore and determine $\mathbf{\Phi}$, subsequently obtaining $\mathbf{\hat{F}}$ through equation \eqref{finalfhat}. In contrast, ORESOU employs DDPG to explore both $\mathbf{\overline{\Phi}}$ and $\mathbf{\hat{F}}$. Consequently, the action space for IC encompasses only $\mathbf{\Phi}$, while ORESOU's action space encompasses both $\mathbf{\overline{\Phi}}$ and $\mathbf{\hat{F}}$. This differentiation captures the key distinction in the action formulation between the two algorithms. 
In comparison to conventional DDPG, where the action space encompasses $\mathbf{\Phi}$ and $\mathbf{\hat{F}}$, IC simplifies the action space by disregarding $\mathbf{\hat{F}}$. On the other hand, ORESOU streamlines the action space by employing binary $\mathbf{\overline{\Phi}}$ in place of the continuous $\mathbf{\Phi}$.

\section{Numerical Simulations}
In this section, we conduct a comprehensive performance evaluation of IC and ORESOU algorithms, both implemented using the DDPG algorithm. Simultaneously, to highlight the advantages of our proposed methodologies, we compare them against two baseline approaches: a conventional DRL method and a random search strategy.

The DRL method involves utilizing DDPG to explore the continuous phase shift configuration $\mathbf{\Phi}$ and the BS beamforming matrix $\mathbf{F}$. Conversely, the random search approach generates $\mathbf{\Phi}$ and $\mathbf{F}$ randomly, ensuring adherence to the conditions specified in equation \eqref{conditions}. For the DDPG implementation, we adopt the well-established implementation available on GitHub \cite{saglam2021}. Key system setting parameters are detailed in Table \ref{systempara}, and we focus on a simplified scenario featuring 2 RISs and 2 UAVs.

\begin{table}[htb]
\caption{PARAMETER VALUES USED FOR THE SYSTEM MODEL}
    \centering
    \begin{tabular}{|c|c|}
    \hline
       Parameter  & Value \\\hline
        BS coordinates (m) & (0,0,2)\\\hline
        RIS coordinates\tnote{1} (m) & (-2.5,8,0),(2.5,0,0)\\\hline
        UAV coordinates\tnote{1} (m) & (-3,10,6),(2,6,10)\\\hline
        $\beta$ &0\\\hline
        $N_{\text{B}}$ & 4\\\hline
        $N_{\text{U}}$ & 2 \\\hline
        $N_{\text{R}}$ & 2 \\\hline
        $N$ & 16 \\\hline
        $f_c$ & 28 GHz \\\hline
        $\kappa_1$,$\kappa_2$&30 dB\\\hline
        $\sigma_z^2$ &-100 dBm\\\hline
    \end{tabular}
     \label{systempara}
\end{table}

\begin{figure}[htb!]
\centering
\begin{subfigure}{.45\textwidth}
  \centering\includegraphics[width=0.8\linewidth]{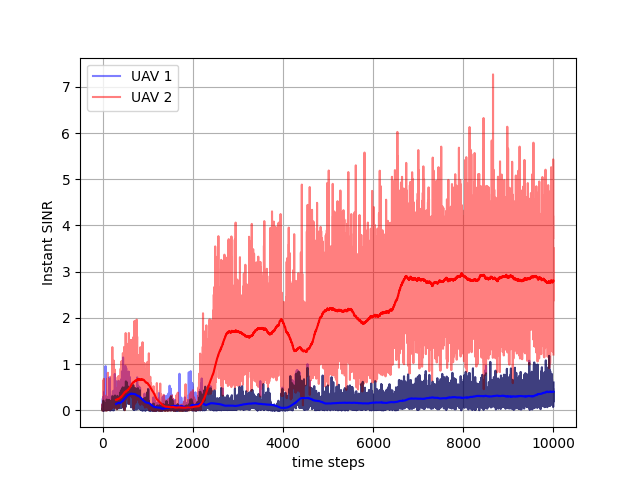}
\caption{Objective: Maximize the sum rate.}
\label{obj_sr}
\end{subfigure}
\begin{subfigure}{.45\textwidth}
\centering
\includegraphics[width=0.8\linewidth]{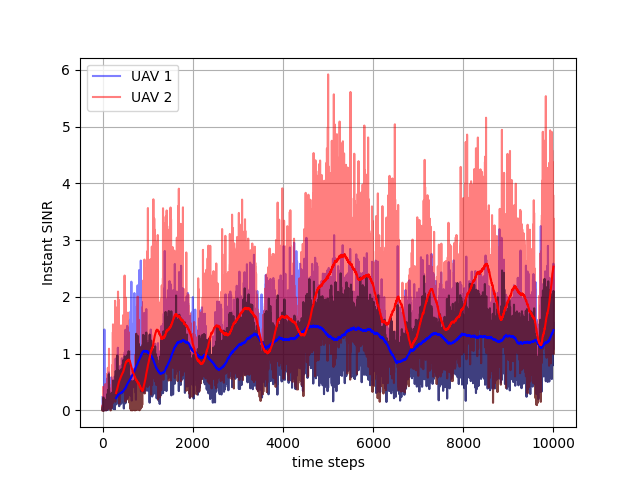}
\caption{Objective: Maximize the minimum SINR.}
\label{obj_msinr}
\end{subfigure}
\caption{Instant and average SINRs of two UAVs using two objectives (rolling average window = 300 time steps).}
\end{figure}

\begin{figure}[htb!]
\centering
  \centering
  \includegraphics[width=0.7\linewidth]{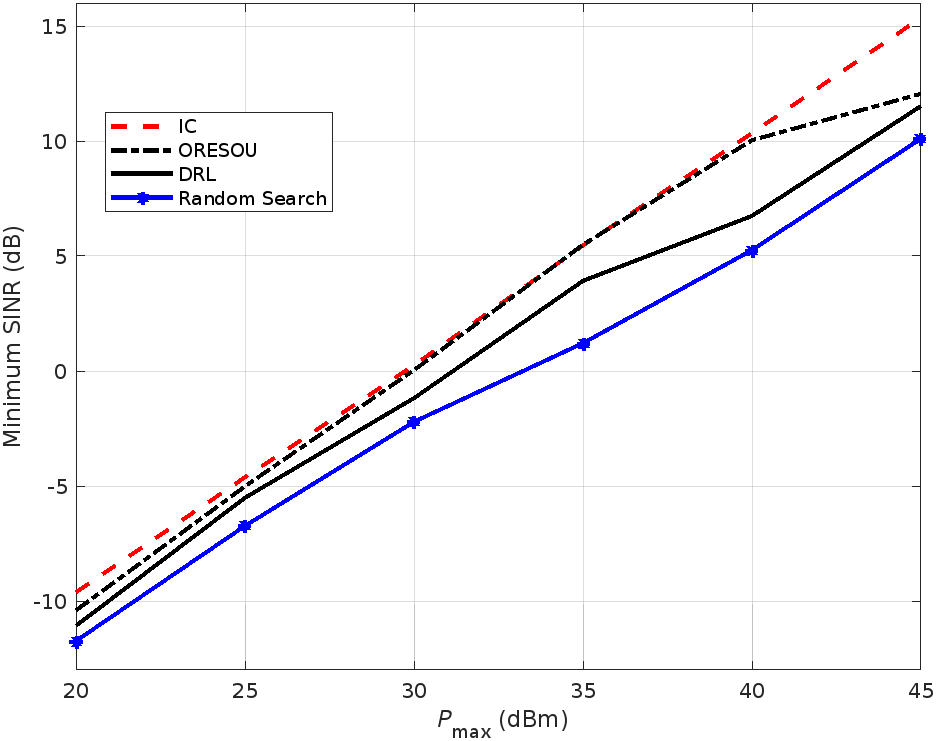}
\caption{Minimum SINR versus $P_{\max}$.}
\label{pmax_sinr}
\end{figure}

\begin{table}[htb!]
\caption{Comparison between one-bit and continuous RIS with $P_{\max}=45$ dBm}
\centering
\begin{tabular}{|c|c|c|c|c|}
\hline
Minimum SINR (dB)&\multicolumn{1}{l|}{IC} & \multicolumn{1}{c|}{ORESOU} & \multicolumn{1}{l|}{DRL} & \multicolumn{1}{l|}{Random Search} \\ \hline
One-bit& 11.63 & 10.167   & 9.79  & 4.058  \\ \hline
Continuous & 15.36 & 12.05   & 11.52 & 10.10 \\ \hline
\multicolumn{1}{|c|}{Difference} & 3.73 & 1.89   & 1.73 & 6.05  \\ \hline
\end{tabular}\label{onebit}
\end{table}
We first investigate the issue caused by setting the objective function as maximizing the sum rate. In this experiment, we use DDPG to explore $\mathbf{\Phi}$ and $\mathbf{F}$ to achieve the maximum sum rate with $P_{\max}=35$ dBm. The recorded instant and average SINRs of two UAVs at each time step are shown in Fig. \ref{obj_sr}. Note that the average SINR is computed by a rolling window of 300 time steps. We can see that the SINR of `UAV 1' tends to very low while the SINR of `UAV 2' becomes very high. In reference to Table \ref{systempara}, we can observe that `UAV 2' experiences less propagation loss from the 2 RISs compared to `UAV 1'. Consequently, it can be deduced that `UAV 2' possesses a superior channel condition. We can conclude that DRL tends to assign most of the resources to the UAV with the best channel condition to achieve the largest SINR and the maximum sum rate. When formulating the objective function to maximize the minimum SINR, the issue of SINR imbalance between the two UAVs is negligible, as illustrated in Fig. \ref{obj_msinr}. The discrepancy in both instant and average SINRs for these two UAVs is much smaller, emphasizing the consideration for fairness among different UAVs. Therefore, we opt for maximizing the minimum SINR as our objective function.

In the evaluation of the IC, ORESOU, DRL, and random search algorithms, we analyze their best minimum SINRs for `UAV 1' and `UAV 2' over 10,000 iterations with a varying $P_{\max}$ ranging from 20 to 45 dBm, as depicted in Fig. \ref{pmax_sinr}. Notably, we observe an increasing trend in minimum SINR with the increment of $P_{\max}$. Among the algorithms, IC consistently outperforms the others, exhibiting superior performance in terms of minimum SINR. ORESOU closely follows IC in performance, demonstrating a slightly lower but significantly improved performance compared to the two baseline approaches, DRL, and random search. In comparison to DRL, IC demonstrates an enhancement ranging from 1.5 dB (for $P_{\max} =20$ dBm) to 4 dB ($P_{\max}=45$ dBm), while ORESOU exhibits improvement spanning from 0.5 dB ($P_{\max} =25$ dBm) to 3.3 dB ($P_{\max}=40$ dBm). 

In the context of practical engineering implementation, the complexity associated with continuous phase configuration in RIS systems has led researchers to explore the viability of employing a simplified one-bit phase configuration. In this approach, the phase shift of each RIS element can only take on two values: 0 or $\pi$, significantly reducing the implementation complexity. In our evaluation, we compared the performance of our proposed algorithms and baseline approaches under the one-bit RIS configuration with a maximum power constraint of $P_{\max}=45$ dBm. The results, as presented in Table \ref{onebit}, reveal that adopting a one-bit RIS configuration can lead to substantial performance degradation across all four algorithms. Specifically, the Minimum SINR of IC experiences a significant 3.73 dB reduction. Both ORESOU and DRL algorithms show a similar and moderate reduction of about 1.8 dB. In contrast, the random search approach exhibits the most substantial performance degradation, with a fall of over 6 dB. Despite these challenges, IC and ORESOU still outperform the DRL and random search approaches in this one-bit RIS setting. The consistent superiority of IC and the competitive performance of ORESOU underscore the efficacy of our proposed algorithms in optimizing resource allocation and tackling interference mitigation issues.


\section{{Conclusions}}\label{sec:conclusion}

In this paper, we have focused on fostering fairness among various UAVs within a multi-RIS-augmented 5G MIMO network, aiming for seamless 3D connectivity. Our specific goal is to maximize the minimum SINR by concurrently optimizing transmit beamforming vectors and RIS phase shift parameters. To address the intricate NP-hard non-convex problem, we have employed the cutting-edge DDPG algorithm. Recognizing the challenge posed by the extensive action search space inherent to DRL, we have introduced two novel algorithms, ORESOU and IC. Through comprehensive numerical simulations, we have thoroughly analyzed and compared the performance of ORESOU, IC, DRL, and random search algorithms. Our results have highlighted that IC and ORESOU not only effectively reduce the action space but also significantly enhance performance, particularly in achieving the best minimum SINR. These findings verified the efficacy of our proposed algorithms in optimizing the performance of multi-RIS-augmented 5G MIMO networks.



\bibliographystyle{IEEEtran}
\bibliography{IEEEabrv,Ref}
\vspace{12pt}

\end{document}